\documentclass{PoS}

\PoS{PoS(LAT2005)276}

\title{Towards an algebraic approach to the discretisation of fermions. 
}

\ShortTitle{Towards an algebraic approach to the discretisation of fermions. }

\author{\speaker{Vivien de Beauc\'e}\thanks{Work supported by the Japan Society for the Promotion of Science. V. de B would like to thank Noboru Kawamoto for useful discussions.}\\
        Department of Physics, Hokkaido University, Sapporo, 060-0810, Japan \\
        E-mail: \email{debeauce@particle.sci.hokudai.ac.jp}}

\abstract{A discretisation scheme for differential geometry is applied to the problem of constructing lattice actions, the naive and staggered action are thus derived. It is found that after specifying an ansatz for the space of fields, the corresponding lattice action is obtained. The gauging procedure, and the applicability of the method to twisted super-symmetry on a lattice is outlined. Some comments on the QED axial anomaly are made, for the theory in which the lattice projection operator is not inserted.}

\FullConference{XXIIIrd International Symposium on Lattice Field 
Theory\\
		 25-30 July 2005\\
		 Trinity College, Dublin, Ireland}

\begin{document}

Recently, renewed interest has been directed to model building of lattice theories in relation to chiral symmetry \cite{Chandrasekharan:2004cn}. The Ginsparg-Wilson relations have provided an optimal way to break chirality (P. Hasenfratz), even preserving a remnant of the symmetry (L\"uscher). The axial anomaly in QED is reproduced correctly. Also, the Poincar\'e lemma on the lattice may be written, thus, encoding algebraic topology in the formulation. Such considerations suggest the need for a more algebraic formulation of lattice (although not using the non-commutative geometry, see the discussion in \cite{Gockeler:1998bi}).  We argue that differential geometry should also be captured.

Our starting point is a 2D square complex $K$ (the treatment in 4D is almost identical), together with the chain space, i.e linear combinations of vertices, edges, squares (the setup of homology). We start with the Dirac-K\"ahler (DK) equation which is differential geometric. The task is then to discretize the various operations required, which are $d$, the exterior derivative, $\wedge$ the wedge product, $\star$ the Hodge star operator, and the interior product $i_v$. All these operations play an important part in the model. The first lattice transcription of the DK equations, based on chains was done by Becher and Joos \cite{Becher:1982ud}. A central feature of that method, is that although the fermion field is a chain, the coefficients are placed at vertices only (i.e on the lattice), requiring the introduction of lattice translation operators in order to satisfy some of the defining algebraic relations of the operators $(\wedge,\, d, \, \star)$. The Dirac equations arising from the DK equation could not be decoupled, and so one was left with the lattice DK fermion with its multiplicity of "flavors". Recently it was recognized that the Whitney map $W$ of algebraic topology makes for a neater discretization in which there is no need for displacement operators \cite{deBeauce:2003zd}, but the underlying lattice plays no special role, so a direct comparison with lattice actions was not obvious. In the meantime, a solution to the problem of defining a satisfactory discrete interior product was given \cite{deBeauce:2004cw}, the key step is the introduction of a product $\times$ on the space of chains, so that topology is captured in the space of chains as usual, while an element of continuous geometry is captured in the product space, in a consistent way. 

In this talk\footnote{The details will be given in a forthcoming article} we revisit the problem of fermions with this new technique in mind. A new discrete Hodge star will also be proposed. The lattice plays a distinguished role, and arises naturally in the ansatz for the discrete fermion field $\psi^K$. We find: the discretisation satisfies the necessary algebraic relations, the reduction of the Dirac equation is carried out explicitly, and by choosing the ansatz $\psi^K$, $\psi^S$, one finds the naive $\mathcal{S}_F$ and staggered $\mathcal{S}_S$ actions after application of a lattice projection operator $\mathcal{P}$ to the "true action" $\mathcal{S}$, then we discuss briefly the question of the chiral anomaly for the theory based on $\mathcal{S}$. At the end, we discuss gauging and sketch the applicability to SUSY.

\section{Mathematical tools}
As announced, a central role is given to the Whitney map $W$. It associates a differential form to a chain, so that a vertex is mapped to a zero-form, an edge to a one form and a square to a two-form. In order to express the resulting forms, we need coordinates, these are given by the hypercubic generalization of the standard simplex. In each $2D$ elementary cell $\sigma_{i, (2)}$ of the complex $K$, we introduce local coordinates $(x,y)$, in the interval $[0,1]\times[0,1]$. For the present purpose, the main point about the Whitney forms is that given a face $\sigma_{j, (p)}$ of a given square $\sigma_{i, (2)}$, the Whitney form $W(\sigma_{j, (2)})$ is defined over the entire coordinate region associated to $\sigma_{i, (2)}$. For example\footnote{For the square $[0123]$, $[0]$ has coordinates $(0,0)$, $[1]$ is $(1,0)$, $[2]$ is $(1,1)$ and $[3]$ is $(0,1)$.}   $W([01]) = (1-y) dx$ is defined over the entire interval $[0,1]\times [0,1]$. This simple fact is conceptually important, the cells of highest dimensionality serve as local open sets in which operations can be defined. The discrete version of the exterior derivative $d^K$ is such that 
\begin{equation}
dW = W d^K,
\end{equation}
which means that if the space of Whitney forms is considered, the discrete operation is exact, at every point in the interval $[0,1]\times [0,1]$ corresponding to the square $\sigma_{i, (2)}$.

For the discrete Hodge star, the commonly used discrete operator is only approximate. A dual complex $L$ is introduced, defined by joining the barycenters of edges and squares of the original complex $K$ (image of $K$ under $\star^K$), of course, the resulting cells $\sigma_{i,(2)}^L$ overlap with four different squares $\sigma_{j,(2)}^K$, so there is no way the operation can be exact. Nevertheless, by using a dual lattice, the co-homology is correctly captured and the Poincar\'e duality $H^{(p)}(M) \cong H^{(2 -p)}(M)$ indeed holds. A discrete wedge is introduced ($N(p,q)$ is a normalization)
\begin{equation}
\sigma_{i, (p)} \wedge^K \sigma_{j, (q)} =  N(p,q)\sum_k\sigma_{k, (p+q)} \int_{\sigma_{k, (p+q)}} W(\sigma_{i, (p)}) \wedge W(\sigma_{j , (q)}). 
\end{equation}
The discrete wedge is non-associative, for the simplicial version it is due to the presence of numerical factors, for the hyper-cubic complex, it is dependent on the chains having overlapping support, since if $\sigma_{i, (p)} \cap \sigma_{j, (q)} \neq \emptyset$ then $\sigma_{i, (p)} \wedge^K \sigma_{i, (q)} = 0$ in contrast Whitney forms always have overlapping support in the interval $[0,1] \times [0,1]$ corresponding to $\sigma_{i, (2)}$. 

We now supply this scheme with a contraction operation \cite{deBeauce:2004cw}. It is defined using a Cartesian product on chains with a particular interpretation, it is
\begin{equation}
i_{v^K} \sigma_{(r)} \doteq v^K \times \eta_{v^K} ( \sigma_{(r)}),
\end{equation} 
where both the vector and contracted chain $\eta_{v^K} ( \sigma_{(r)})$ are mapped to Whitney forms but the chain on the left of $\times$ is interpreted as a function coefficient. The Whitney map is then modified to 
\begin{equation}
\bar{W} ( \sigma \times \beta) \doteq \varphi^{(0)}( W(\sigma)) \wedge W(\beta).
\end{equation}
The map $\varphi^{(0)}$ deletes the form part of $W(\sigma)$. This product $\times$ is compatible with the other operations. A key property is that the contraction is exact and the product is local in the sense that both $W(\sigma)$ and $W(\beta)$ should have overlapping support to give a non-zero result. 

To start describing what is new in this talk, we propose to replace the discrete Hodge star which maps one complex to its dual by one which uses the product space $\times$ instead. Let
\begin{equation}
\label{hodge}
\star^K \sigma_{i, (p)} \doteq \sigma_{i, (p)} \times \sigma^{d(i,p)},
\end{equation}
where $\sigma^{d(i,p)}$ maps under $W$ to one of the globally constant forms\footnote{For example $W(\sigma^{\mu}) = dx^{\mu}$.} $\{1, \; dx^1, \; dx^2,\; dx^1 \wedge dx^2\}$.
The new discrete $\star$ is exact, and does not require a double complex.
Previously, the presence of the dual lattice $L$ doubled the degrees of freedom, a form of species doubling\footnote{also the new $\star$ playing the role of $\gamma^5$ projects correctly to the lattice under $\mathcal{P}$, below.}. Finally, the contraction operation may be used to define a discretized Clifford product ($f$ and $g$ are chain degree dependent)
\begin{equation}
\sigma_{i, (p)} \vee \sigma_{j , (q)} \doteq \sum_{p} (-1)^{f} ((-1)^{g}i_{\partial_{\mu_1}}^K \ldots  i_{\partial_{\mu_p}}^K \sigma_{i, (p)} ) \wedge^K i_{\partial_{\mu_1}}^K \ldots i_{\partial_{\mu_p}}^K \sigma_{j, (q)}.
\end{equation}
The discrete wedge is approximate, but it is exact when one of the two chains maps to a constant form, this is the case with the chiral projection operators $P^{(b)}$.  
\section{Derivation of the lattice actions}
Let us start by writing down the natural action, 
\begin{equation}
\mathcal{S} = \int_M W( \bar{\phi}^K)\wedge \star W( (d - \delta) \phi^K),  
\end{equation}
having then specified the inner product, what remains is to find how to parametrize $\phi^K$. In the continuum theory, the field is an in-homogeneous form i.e
\begin{equation}
\phi^K = \phi + \phi_{1} dx^1 + \phi_2 dx^2 + \phi_{12} dx^1 \wedge dx^2.
\end{equation}
By application of the chiral projection operator $P^{(b)}$, the equation is reduced to the Dirac equation, the two-component spinor $\psi^{(1)} = \psi \vee P^{(1)}$ is then
\begin{equation}
\label{psi}
\psi^{(1)} = \phi_1 ( 1- i dx^1 \wedge dx^2) + \phi_2 ( dx^1 - i dx^2).
\end{equation}
The projection can be carried out in the present model as we will see, and the discrete DK operator is now exact i.e
\begin{equation}
\bar{W} ( d^K - \delta^K) = (d - \delta) \bar{W}.
\end{equation}
The ansatz for the discrete fermion field is the key to deriving the lattice actions. We note that with the new discrete Hodge star, it is mandatory to use the product $\times$ for the basis of fields, since otherwise it can be shown that $\delta \sigma = 0$ for any chain $\sigma$. The ansatz then is the following discrete version of Eq. \ref{psi}, where the coefficients $\lambda_{(i,j)}$ are placed on the lattice (labelled by a pair of integers $(i,j)$), and
\begin{equation}
\label{psik}
\psi^K = \sum_{(i,j)} \lambda^1_{(i,j)} [(i,j)] - i (\sum_{(i,j)} \lambda^1_{(i,j)} [(i,j)]) \times \sigma^{12} + \lambda^2_{(i,j)} [(i,j)] \times( \sigma^1 - i \sigma^2), 
\end{equation}
we then verify it is invariant under $P^{(1)}$ and substitute this ansatz into the DK equation
\begin{equation}
(d^K - \delta^K) \psi^K = 0.
\end{equation}
This leads to four equations, of the form $(\sum_k \lambda_k \sigma_k) \times \sigma^{d}$. The chains on the left of $\times$ form two pairs of identical equations. The two independent equations are
\begin{eqnarray}
\sum_{(i,j)} ( \lambda^2_{(i+1,j)} - \lambda^2_{(i,j)}) [(i,j)(i+1,j)] - i ( \lambda^2_{(i, j+1)} - \lambda^2_{(i,j)}) [(i,j)(i, j+1)] &=& 0 \\
\sum_{(i,j)} ( \lambda^1_{(i+1,j)} - \lambda^1_{(i,j)} ) [(i,j)(i+1,j)] - i (\lambda^1_{(i, j+1)} - \lambda^1_{(i,j)})[(i,j)(i,j+1)] &=& 0
\end{eqnarray}
These equations, after application of $\bar{W}$, appear to be approximate in that they are functions of $1, x^1, x^2$ and have no non-trivial solutions in the coefficients $\lambda_{(i,j)}$. Nevertheless we can now connect to the lattice formulation, by noting that if we evaluate $\bar{W} ( d^K - \delta^K) \psi^K$ at a given vertex $[(i,j)]$, we find precisely the naive $2D$ Dirac equation on the lattice
\begin{equation}
\sigma_{\mu} \partial_{\mu}^S \psi^K_{(i,j)} = 0, 
\end{equation}
where $\partial_{\mu}^S$ is the symmetric difference operator. The lattice action can then be derived by inserting the projection operator $\mathcal{P} = \sum_{(i,j)} \delta(x - x_{(i,j)})$ in the action ($x_{(i,j)}$ are the coordinates of $[(i,j)]$), i.e 
\begin{equation}
\mathcal{S}_F = \int_M \mathcal{P} W(\bar{\psi}^K) \wedge \star W (  (d^K - \delta^K) \psi^K)
\end{equation}
where $\psi^K$ is the ansatz Eq.\ref{psik}. For the staggered fermion, we modify the ansatz to $\psi^B_{S}$ below. Of course one needs to introduce a double spacing for spinors, this may be done by introducing a dual lattice (same as $L$ but with the new Hodge star) with its own collection of operations $(\wedge, d , \star, i_v)^L$, defined as we did in $K$. The ansatz is then (the Kronecker delta identify even and odd sites):
\begin{eqnarray}
\label{stag}
\psi^B_{S} &=& \sum_{(i,j)} \delta_{i\,(2) ,0}\delta_{j \, (2), 0} \lambda^1_{(i,j)} [(i,j)] - i (\sum_{(i,j)} \delta_{i\, (2), 1}\delta_{j \, (2) , 1} \lambda^1_{(i,j)} [(i,j)]) \times \sigma^{12} \\
&+& \lambda^2_{(i,j)} [(i,j)] \times(\delta_{i\, (2), 1}\delta_{j\, (2), 0} \sigma^1 - i\delta_{i\, (2),0}\delta_{j\, (2), 1} \sigma^2), 
\end{eqnarray}
after subdividing the cells into the collection of finer ones $B = K \cap L$, one can write the action $\mathcal{S}(\bar{\psi}^B_S, \, \psi^B_{S})$ and after application of the lattice projection operator $\mathcal{P}$ one obtains the staggered fermion action.

Let us now take a different viewpoint, we do not insert the projection operator $\mathcal{P}$ in the action $\mathcal{S}$, and we evaluate $\mathcal{S}(\bar{\psi}^K, \psi^K)$ to get a matrix $M$ such that
\begin{equation}
\mathcal{S} ( \bar{\psi}^K, \psi^K) = \sum_{(i,j)} \bar{\psi}_{(i,j)} M \psi_{(k,l)}, 
\end{equation}
it contains more couplings than the projected action $\mathcal{P} \mathcal{S}$, namely diagonally opposite vertices in a square are coupled. This still respects the locality, and is chiral. On can add a Laplacian term 
\begin{equation}
\bar{\psi} \wedge \star (d \delta + \delta d) \psi 
\end{equation} 
which breaks chirality and then project onto the lattice to a Wilson term. 
It is important to note that the lattice theory defined with the operator $M$ does not have the same symmetries as the action $\mathcal{S}$ which has continuous symmetries. This brings the possibility of using the action $\mathcal{S}$ with the lattice parametrization of the field space $\psi^K$. In this case one can write explicitly the chiral current in differential geometric language. Since then chiral symmetry is not broken at the classical level, there is a strong suggestion that the anomaly may be derived between lattice and continuum in the framework presented here (there is no momentum cut-off). 

We now discuss gauging. The gauge field is a one-form, so it is associated to edges as
\begin{equation}
A^K = \sum_{(i,j)} A_{(i,j)(i+1,j)} [(i,j)(i+1,j)] + A_{(i,j)(i, j+1)} [(i,j)(i, j+1)], 
\end{equation}
this makes the parallel with the link variable formulation easy to obtain. We consider the non-compact gauging
\begin{equation}
\mathcal{S}_G = \int_M W (\bar{\psi}^K) \wedge \star W ( (d^K - \delta^K + A^K \vee^K)\psi^K), 
\end{equation}
after inserting the projection operator, and identifying the leading terms in the expansion of the link variable $U = e^{i A}$, we find (showing only $x$-axis link variables, $\hat{k}$ are lattice unit vectors)
\begin{equation}
\mathcal{P} \mathcal{S}_G = \sum_{(i,j)} \bar{\psi}^K_{(i,j)} U \psi^K_{(i+1,j)} + \sum_{(i,j)} \bar{\psi}^K_{(i,j)} \psi^K_{(i,j)}\sum_k A_{(i,j),(i,j) + \hat{k})},  
\end{equation}
the second term being not gauge invariant, we must impose a path ordering to prevent such term. It is then clear that the slightly non-local nature of the wedge product (within a cell) enables one to extract the link variable. Note that the latter cannot be used as a Wilson type term since it is not gauge invariant. An alternative non-compact treatment is to do the minimal coupling without using the wedge product but rather using the product $\times$. Such approach also enables one to consider pure gauge theory and one finds that $d_A^2 = F$. 

We conclude with our prescription for lattice SUSY. A link between the DK fermion and extended twisted SUSY (which is a topological theory) has been made \cite{Kawamoto:1999zn}. The corresponding lattice theory is link variable based, and at an algebraic level is described by non-commutative geometry \cite{D'Adda:2005zk}. Our proposal also works at the algebraic level: the super-parameters are not discretized (e.g Wess-Zumino action), the partial derivatives are replaced with $\partial^{\mu K} \doteq (d^{\mu} \sigma) \times \sigma^{0}$, where $d^{\mu}$ is the $\mu$-axis component of $d^K$. The DK part of the twisted action is as defined above, including both chiral projections. This corresponds to $D= N =2$. For example,
\begin{equation}
Q = \frac{\partial}{\partial \theta} + \frac{i}{2} \theta^{\mu} \partial_{\mu}  \mapsto \frac{\partial}{\partial \theta} + \frac{i}{2} \theta^{\mu} \partial_{\mu}^K.
\end{equation}
One immediately verifies that the algebra
\begin{equation}
\{ Q, \, Q^{\mu} \} = i \partial_{\mu}, \; \{ \tilde{Q}, \, Q^{\mu} \} = - i \epsilon_{\mu \nu} \partial^{\nu},
 \end{equation}
is realized on the field $\psi_S = \psi_S^K + \psi_S^L$, where the flavor is identified as a twisted SUSY index.

\end{document}